\documentclass[conference]{IEEEtran}
\IEEEoverridecommandlockouts
\usepackage{cite}
\usepackage{amsmath,amssymb,amsfonts}
\usepackage{graphicx}
\usepackage{textcomp}
\usepackage{xcolor}
\usepackage{optidef}
\usepackage[mathscr]{euscript}
\def\BibTeX{{\rm B\kern-.05em{\sc i\kern-.025em b}\kern-.08em
    T\kern-.1667em\lower.7ex\hbox{E}\kern-.125emX}}

\def\SINR{\mathrm{SINR}}
\def\E{\mathbb{E}}
\DeclareMathOperator*{\minimize}{minimize}

\usepackage{algorithm}
\usepackage{algpseudocode}
\renewcommand{\baselinestretch}{0.92}
\begin{document}

\title{Towards URLLC with Proactive HARQ Adaptation \\

}
\author{\IEEEauthorblockN{Lam Ngoc Dinh\IEEEauthorrefmark{1}, Ibtissam Labriji\IEEEauthorrefmark{2}, Mickael Maman\IEEEauthorrefmark{1} and Emilio Calvanese Strinati\IEEEauthorrefmark{1},}

\IEEEauthorblockA{ 
    \IEEEauthorrefmark{1} CEA-Leti, Université Grenoble Alpes, F-38000 Grenoble, France \\
    \IEEEauthorrefmark{2} Renault Software Labs, 06560 Valbonne, France \\
\{ngoc-lam.dinh, mickael.maman, emilio.calvanese-strinati\}@cea.fr, ibtissam.labriji@renault.com\\} 
}

\maketitle

\begin{abstract}
In this work, we propose a dynamic decision maker algorithm to improve the proactive HARQ protocol for beyond 5G networks. 
Based on Lyapunov stochastic optimization, our adaptation control framework dynamically selects the number of proactive retransmissions for intermittent URLLC traffic scenarios under time-varying channel conditions without requiring any prior knowledge associated with this stochastic process. It then better exploits the trade-off between Radio Access Network (RAN) latency, reliability and resource efficiency, which is still limited in its realization on current HARQ designs. We then evaluate the performance of several HARQ strategies and show that our proposal further improves latency over the reactive regime without affecting the resource efficiency such as fixed proactive retransmission while maintaining target reliability.
\end{abstract}


\section{Introduction}
The 5G and beyond network enables the exploitation of new emerging use cases, such as Ultra-Reliable Low Latency Communication (URLLC) which is applicable for mission critical services, remote surgery and wireless factory control. However, ensuring URLLC in wireless communication is a challenging task as it requires the simultaneous enhancement of low latency and high reliability (e.g., industrial control demands a packet error rate of $10^{-5}$ with a latency less than 1 $ms$). 
Under such stringent requirement, the retransmission regime in Hybrid Automatic Repeat reQuest (HARQ) is inadequate since each Round Trip Time (RTT) could take several time slots  \cite{kimImplementationDownlinkAsynchronous2012} for the corrupted packet to be retransmitted again and thus the latency requirement is no longer met if many retransmissions (RTXs) are required.

In order to tackle with the problem of long RTTs for mission critical applications,  a K-repetition regime \cite{leOverviewPhysicalLayer2021a} and a proactive regime with early termination \cite{liuAnalyzingGrantFreeAccess2021a} have been proposed. By performing such fixed repetition of a packet multiple times, regardless of the RTX success or failure, one can opportunistically decode the packet at the receiver in a shorter time at the expense of inefficient resource utilization \cite{jacobsenSystemLevelAnalysis2017a}.  
To cope with the over-estimation of packet repetition, Le et al. \cite{leOptimalReservedResources2019} proposed a reserved resource scheme to guarantee that each HARQ process can perform an adequate number of repetitions under a configured period. However, in a time-varying radio channel, the lack of an adaptive packet RTX strategy evidently degrades the system performance since the number of RTXs is applied imprecisely and both resource utilization and latency are negatively impacted. In the current literature, the adaptation of the HARQ strategy can be achieved in several ways such as adapting the modulation and coding scheme \cite{pfletschingerAdaptiveHARQNonBinary2014}, the transmission power \cite{jabiEnergyEfficiencyAdaptive2016} and the maximum number of RTXs \cite{mukhtarContentawareOccupancybasedHybrid2016}. By optimally tuning these parameters under dynamic channel conditions, the performance of adaptive HARQ can be substantially improved. 


Considering the benefits of proactive HARQ and the adaptation strategy to dynamic channel conditions, in this paper, we proposed an adaptation control algorithm to improve proactive HARQ. In addition, we evaluate our design in an intermittent traffic scenario that is close to the use cases of URLLC in 5G. To the best of our knowledge, there is still limited amount of research that considers such dynamics in the results.  In \cite{Han21}, they proposed a Closed-Loop ARQ protocol that dynamically re-allocates
the remaining resource between uplink and downlink slots upon the result of last uplink transmission.
In contrast to the current state-of-the-art, our proposition will robustly select the number of RTXs sufficient to maintain a good level of resource efficiency and ensure a shorter packet delivery time. To design such an optimal control action, knowledge of stochastic processes such as traffic behaviour and instantaneous channel quality is required. However the acquisition of such information is usually difficult and their behaviours are unpredictable, which is an obstacle to design a feasible optimization algorithm. Fortunately, the framework of so-called Lyapunov stochastic optimization does not require the prior probabilities associated with these processes \cite{neelyStochasticNetworkOptimization2010}. Furthermore, given the compatibility of the time-slot system in the 5G New Radio (NR), the Lyapunov optimization framework becomes a promising candidate to deal with our problem. The optimal performance of the balance between shorter packet delivery and better resource usage is then achieved by fine-tuning the  \textit{V} parameter of the Lyapunov framework.    

Based on rich information perceived prior to the control action to be taken, such as the current state of the queue backlog, our objective is to minimize the long-term average of total resource allocation needed for each packet, subject to the stability of the queue and the reduction of the risk of bad decisions. To formalize our problem, we presented the dual objectives of resource allocation and queue stability in \textit{drift-plus-penalty problem}, and then introduced a \textit{virtual queue} whose evolution is associated with the risk of taking an inadequate proactive retransmission.

The remainder of the paper is organized as follows: Section II presents a detailed scenario setup where many physical (PHY) and Medium Access Control (MAC) layers assumption are described in our work and is followed by problem formulation. Then, Section III details our designed algorithm. In Section IV, the simulation results demonstrate the performance of the system and Section V concludes the paper.

\section{System Model}
\subsection{Scenario Description}

The network contains 1 gNb and 1 UE. During the simulation, the UE node moves around  the gNB and varies its distance to it. Downlink (DL) traffic is assumed to be generated by a remote host located near the gNb, in which the variable arrival rate and packet length follow an exponential distribution. The application packets are then queued in the transmission buffer $Q_1(t)$ of the Radio Link Control (RLC) layer.  After completing the scheduler operation at MAC layer, the gNb prepares a transport block (TB) whose data is extracted from $Q_1(t)$ and sends it over the air. At the same time, the scheduler keeps a copy of this TB and associates it with an identifier to construct an identical HARQ process \cite{3gppMediumAccessControl2019}, which will give the receiver information about the processed data. In order to demonstrate the state of ongoing HARQ processes that are not yet decoded at the UE side, we defined a $Q_2(t)$ that contains them. Due to the dynamic nature of not only the traffic model but also the channel behaviour, the lengths of $Q_1(t)$ and $Q_2(t)$ are influenced and can be considered as random variables. The state of $Q_1(t)$ and $Q_2(t)$ demonstrated a two-stage queuing system whose length should be minimized.  

\begin{figure}[htp!]
\centering
\includegraphics[width=0.46\textwidth]{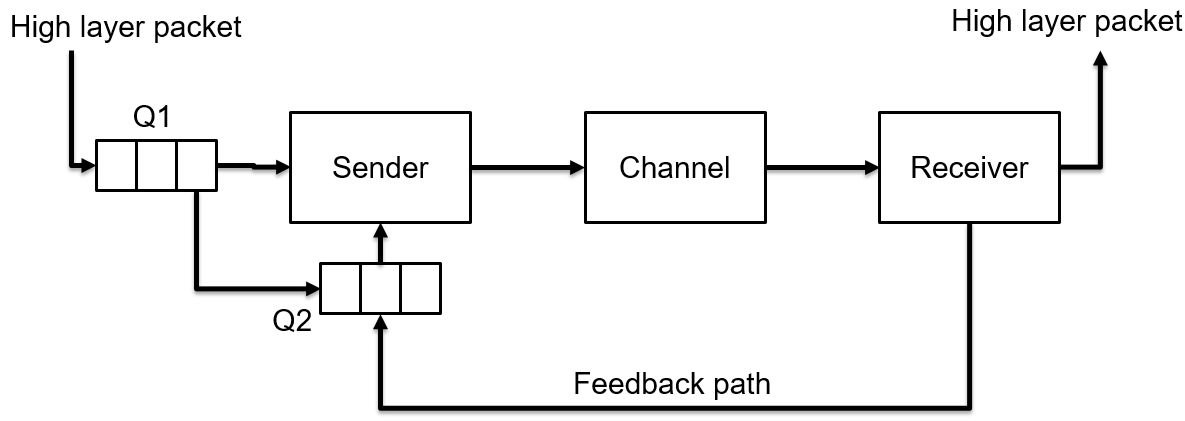}
\caption{System model}
\label{fig:sm}
\end{figure}
To calculate the TB size, the TDMA-based scheduler takes into account the number of OFDM symbols allocated to user in a single time slot ($1 \leq N_{OFDM} \leq 12$), as well as the choice of the Modulation and Coding Scheme (MCS) $m$, the numerology $Num$, the bandwidth $BW$, as follows:
\begin{equation}\label{tbs}
    \mathrm{TB} = \frac{N_{OFDM} \times BW}{8\times\mathrm{SCS}(Num)} \times M(m) \times CR (m)
\end{equation}
where $\mathrm{SCS}(Num)$ is the sub-carrier spacing at the numerology $Num$, $M(m)$, $CR(m)$ are the modulation order and code rate of the selected MCS $m$, respectively. These values are given in Table 5.1.3.1 of 3GPP TS 38.214 \cite{3gppPhysicalLayerProcedures}.

Concerning the improvement of the diversity order of the communication, $U_{tx} = M_{BS} \times N_{BS} $ and $S_{rx}=M_{UE} \times N_{UE}$ Uniform Linear Array (ULA)  antennas are installed at the gNB and UE, respectively. The transmission power at the gNb is set to $P_{tx} $ $dBm$.
In order to evaluate the performance of the system in a highly changing environment, we modeled the Indoor Factory- sparse clutter scenario (InF-SL) according to the 3GPP technical report \cite{3gppTr382019}.
This model clarifies the channel behaviours based on antenna modeling, slow-fading factors such as shadowing, propagation loss and fast-fading factors due to multi path. Initially, the propagation loss is calculated based on whether the path is in visibility or not (LOS or NLOS) and can be derived:
\begin{equation}\label{PL_LOS}
      PL_{LOS}(d) = 31.84 + 21.50\times log_{10}(d) + 19.00\times log_{10}(f_c)
\end{equation}
\begin{equation}\label{PL_NLOS}
      PL_{NLOS}(d) = 33 + 25.50\times log_{10}(d) + 20.00\times log_{10}(f_c)
\end{equation}
where $d$ is the instantaneous 3D distance between the UE and gNb in meter and $f_c$ is the center frequency in GHz.

According to the channel statistics, the probability of LOS communication $Pr_{LOS}$ between the transmitter and receiver which is based on the  distance $d$ and the scenario parameter $k=-\frac{d_{clutter}}{ln(1-r)}$ can be derived as follows. 
\begin{equation} \label{Plos}
    Pr_{LOS} = e^{-\frac{d}{k}}
\end{equation}
where $d_{clutter} = 10$ is the typical clutter size in meter and $r=0.3$ is the clutter density. In this study, we consider the UE mobility with respect to the gNb, so the shadowing effect plays an important role in the channel status and it has been modelled as in \cite{itu-rPredictionTimeSpatial2019}.

\begin{figure}[htp!]
\centering
\includegraphics[width=0.46\textwidth]{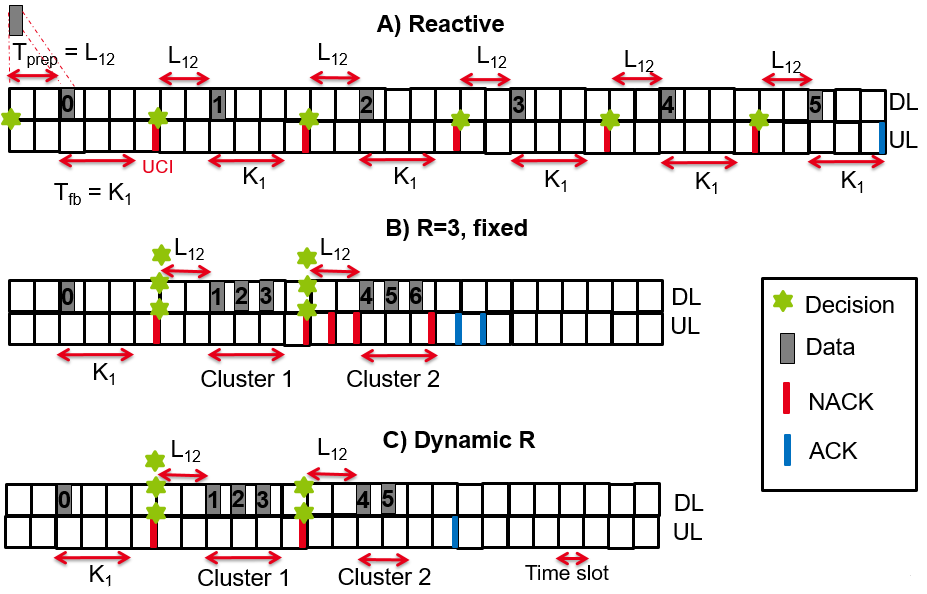}
\caption{Different DL HARQ procedures: (A) classical, (B) fixed R=3 parallel RTXs and (C) Dynamic and proactive RTXs.}
\label{fig:harq}
\end{figure}

To better understand the benefits of good decision making for proactive HARQ in improving Radio Access Network (RAN) latency and resource efficiency, Figure \ref{fig:harq} shows the schemes of \textit{(A)} classical reactive HARQ procedure, \textit{(B)} fixed repetitions of 3 RTXs and \textit{(C)} dynamic redundancy applied to the number of RTXs.

 In Figure \ref{fig:harq}A, a delay $L_{12}$ is introduced to demonstrate the TB preparation time from the gNb scheduler to the antenna. Then, a feedback will be encoded together with an Uplink Control Information (UCI) message and sent back to the gNB after $T_{fb}=K_1$  slot(s), thus illustrating the processing time at the UE. In 5G NR standard, this processing time reflects a delay between the reception of the UL grant in the DL (PDCCH) and the transmission of the corresponding UL data (Physical Uplink Shared Channel (PUSCH)) \cite{patricielloImpactNRScheduling2019}. Afterwards, the gNB has the information about the corrupted HARQ process on the UE side and decides to retransmit the erroneous TB after $L_{12}$ slots. This process continues 5 times until the corrupted TB is successfully decoded by the UE. By doing this, the resource are perfectly utilized, but the latency could be unacceptable for URLLC communications.
Figure \ref{fig:harq}B shows the fixed repetition of corrupted TB with a fixed redundancy level $R=3$. This means that the first cluster which will occupy 3 consecutive slots, will be reserved for retransmissions. If it fails, the second cluster with the same redundancy level will be allocated. As shown in Figure \ref{fig:harq}B, latency is significantly improved at the cost of resource efficiency (RTX6 is useless when the TB has been decoded after 5 retransmissions). 
By dynamically selecting the proactive redundancy level for each cluster, Figure \ref{fig:harq}C shows better performance in terms of reduced latency and improved resource efficiency. In this design, to reduce the control overhead due to multiple feedbacks to the transmitter, we grouped their feedbacks into a single feedback that represents the current proactive retransmission status. 

In this work, a TB is successfully decoded by the receiver if the Signal-to-Noise-plus-Interference-Ratio (SINR) reaches the target $\SINR$.
We consider an incremental redundancy HARQ, thus each RTX contains different coded bits than the previous one. According to \cite{lagenNewRadioPhysical2020}, a singular value of $\SINR^{(r)}$ after $r$ RTXs is computed quantitatively based on a set of received Resource Blocks (RBs) $\omega$, their $\SINR_x^{(r)}$ $(\forall{x \in {\omega}})$ and the previous RTX, i.e. $\SINR_{tb_n}^{(r-1)}$ as follows:
\begin{equation}\label{sinr_ir}
    \SINR_{tb_n}^{(r)} = -\beta\times\mathrm{ln}(\frac{1}{\left | \omega \right |}\times\sum_{x \in \omega}e^{-\frac{\SINR_x^{(r)}+\SINR_{tb_n}^{(r-1)}}{\beta} })
\end{equation}
$\beta$ is a constant depending on the MCS as specified in \cite{lagenNewRadioPhysical2020}.

\subsection{Problem Formulation}
In this work, we limit the number of decisions into \textbf{$c_{max}$} clusters of proactive RTXs. Each cluster $c_j$ $\in \{c_0...\textbf{$c_{max}$}\}$ can include $r_{c_j}$ RTXs. $c_0$ corresponds to the initial TB transmission. The decision maker we designed will dynamically choose the size of each cluster $c_j$ (i.e, $r_{c_j}$) to reduce both RAN latency and resource waste. 
With respect to the resources allocated for proactive RTXs of a $TB_n$, the decision maker selects an element-wise positive resource allocation vector $(r_{n,c_0},r_{n,c_1},...,r_{n,c_{max}})$ that satisfies the following condition:
\begin{equation} \label{cond_r}
    1 \leq \sum _{j} r_{n,c_j} \leq R_{max}
\end{equation} 
where  $R_{max}$ is the maximum number of RTXs for a $TB_n$. In the case where a TB is not decoded at receiver after $R_{max}$ RTXs, the TB that contains many application packets is considered a loss. $r_{min} \leq  r_{n,c_j} \leq r_{max}$ constrains the number of proactive RTXs at cluster $c_j$ not to exceed a value $r_{max}$.   We can define an objective function $f_{obj}$ that is akin to the average resource allocation to be provided for each TB as follows:
\begin{equation}\label{re}
    f_{obj} = \lim _ {N \to \infty} \frac{1}{N} \sum _{n=0} ^{N-1} (\sum _j r_{n,c_j})
\end{equation}

The decision of how to optimally select $r_{c_j}$ is based on various factors, such as the current status of the $Q_1(t),Q_2(t)$, the current cluster index $c_j$ and the current aggregated SINR.


The first decision at time t is to select the queue $Q_i(t)$ $\forall{i\in\left \{1,2  \right \}}$ to proactively (re)transmit the TB over $r_{c_j}$ consecutive time slots. Then, a series of actions $a \in \left \{0,1,2,...  \right \}$ is made at the corresponding action slot $t_a$ in which the proactive RTXs span $r_{a}$ time slots. The time interval ($t_{a},t_{a+1}$) forms the $a-th$ time frame denoted by $F_a$. In order to control at frame $F_a$, which queue among $Q_i(F)$ will be served, we introduced the control variable $\alpha_a$ where $\alpha_a=1$ means serving $Q_1(F_a)$ and  $\alpha_a=0$ means serving $Q_2(F_a)$. Knowing that HARQ processes have a higher priority, $\alpha_a=0$ when $Q_2(t)>0$. 
Then, the queuing dynamic will be given as follows:
\begin{equation}\label{q1}
        Q_1(F_{a+1}) = \max  \{ Q_1(F_a)  -\alpha_a . TB_a^{r_{c_0}}, 0  \} + A_1(F_a) 
\end{equation}
\begin{equation}\label{q2}
\begin{split}
        Q_2(F_{a+1}) = \max  \{ Q_2(F_a)  -& (1-\alpha_a).1_{TB_a^{r_{c_j}}}. TB_a^{r_{c_j}}, 0  \}  \\
        &  + A_2(F_a)
\end{split}
\end{equation}

where $Q_i(F_{a+1})$ are the backlogs of the queue $i$ at time frame $F_{a+1}$. The value $A_1(F_a)=\sum_{t=1}^{F_a}A_1(t)$ represents the total amount of high layer packets that accumulated in $Q_1$ during frame $F_a$. During this frame, an amount of $TB_a^{r_{c_j}}$  that corresponds to the proactive RTX of $TB$ in $r_{c_j}$  slots, will be served. 
The indicator function $1_{TB_a^{r_{c_j}}}$, in Equation \ref{q2}, is equal to 1 if after the $r_{c_j}$ proactive RTXs, $TB$ is successful and is 0, otherwise. If the first transmission of $TB_a^{r_{c_0}}$ is a failure, $A_2(F_a)=TB_a$ will be added as a HARQ process to $Q_2$, otherwise $A_2(F_a)=0$ as the HARQ process of $TB_a^{r_{c_0}}$ will be removed from $Q_2$. According to Little's law, the average delay is related to the queue length. Therefore, the first constraint of our problem is both to minimize the average long-term queue length $\overline{Q_1(F_{a})+Q_2(F_{a})}$ and to make their average rate stable. By definition, a stochastic process Q(t) is stable at the average rate if:
\begin{equation}
    \lim _{t \to \infty} \frac{\mathbb{E}\{Q(t)\}}{t} = 0
\end{equation}

In the following, we present an additional constraint that is associated with the average long term risk. The visible risk in our problem is when the decision maker at cluster $c_{max}$ chooses to serve the HARQ process in $Q_2$ by insufficient allocation of $r_{n,c_{max}}$ to recover the corrupted TB:
\begin{equation}\label{risk}
\begin{split}
    \zeta_n = \mathbb{P} [(\SINR_{tb_n}^{(\sum _{j=0} ^{max} r_{n,c_j})} &\leq \SINR^{t}) \\
    &\mid \SINR_{tb_n}^{(\sum _{j=0} ^{max-1} r_{n,c_j})}]
\end{split}
\end{equation}

where $\SINR_{tb_n}^{(\sum _{j=0} ^{max} r_{n,c_j})}$ is the $\SINR$ of $TB_n$ after $c_{max}$ clusters of RTXs and $\SINR^{t}$ is the target $\SINR$.  

So, the long-term average risk across all TBs is:
\begin{equation}\label{ltrisk}
    \overline{\zeta} = \lim _ {N \to \infty} \frac{1}{N} \sum _{n=0}^{N-1}\zeta _n
\end{equation}

The constraint is then to guarantee the expectation of the long-term risk $\overline{\zeta}$ below a predefined value $\zeta _o$. For that purpose, we introduce a \textit{virtual queue} $Z(t)$ which is incremented each time slot $t$ by $\overline{\zeta}-\zeta _o$ each time $\overline{\zeta}$ exceeds $\zeta _o$.
\begin{equation}
    Z(t+1) = \max \{Z(t) + \overline{\zeta}-\zeta_o ,0\}
\end{equation}
Hence, the constraint of satisfying the long-term risk becomes a stability constraint on the average rate of the queue $Z(t)$. 
\begin{equation} \label{crisk}
    \lim _{t \to \infty} \frac{\mathbb{E}\{Z(t) \}}{t} = 0
\end{equation}

To summarize,  our optimization problem \ref{eq:P1} is to minimize the objective function $f_{obj}$  subject to several constraints:
\begin{align}
    \minimize ~& f_{obj} \tag{$\mathcal{P}_1$} \label{eq:P1}\\
     \text{s.t.}~~ & \lim _{t \to \infty} \frac{\mathbb{E}\{Q_i(t) \}}{t} = 0, & \forall{i \in \{1,2\}}
    \tag{$\mathcal{C}_{1,2}$} \label{eq:C1}\\
    &\lim _{t \to \infty} \frac{\mathbb{E}\{Z(t) \}}{t} = 0, &
    \tag{$\mathcal{C}_3$} \label{eq:C3}\\
    & 1 \leq \sum _j ^{c_{max}}r_{n,c_j} \leq R_{max}, & \tag{$\mathcal{C}_4$} \label{eq:C4}\\
    & r_{min} \leq r_{n,c_j} \leq r_{max} ~~ &  \tag{$\mathcal{C}_5$} \label{eq:C5}
\end{align}

\section{Dynamic Decision Maker Algorithm}
In this section, we propose a dynamic decision maker algorithm based on Lyapunov's optimization tools to solve the optimization problem \ref{eq:P1}. To simplify the design of the sequential decision maker, we assume that \textit{(i)} the first decision is the same as the reactive decision (i.e. only one slot is reserved for the TB transmission) and  
\textit{(ii)} proactive RTXs in the same cluster share the same HARQ feedback. 

First, we define the current state in the slot t as $\Theta(t) = \left(Q_1(t), Q_2(t), Z(t)\right)$ and the Lyapunov function as follows:
\begin{equation}\label{Lf}
    L(\Theta(t)) \triangleq \frac{1}{2}\left[Q_1^2(t) + Q_2^2(t) + Z^2(t)\right]
\end{equation}
Next, we define the one-slot conditional Lyapunov drift $\Delta(\Theta(t))$ representing the expected change of the Lyapunov function over a slot as follows:
\begin{equation}\label{Ld}
    \Delta(\Theta(t)) = \mathbb{E}\{L(\Theta(t+1)) - L(\Theta(t)) \mid \Theta(t) \}
\end{equation}
By minimizing both $\Delta(\Theta(t))$ and $f_{obj}$, we can solve the problem \ref{eq:P1} because the queues are stable in terms of average rate and the objective function is minimized . However, according to \cite{neelyStochasticNetworkOptimization2010}, there is a \textit{performance-delay} trade-off between these dual objective optimizations that can be parameterized by a constant $V$. By setting a large positive value to $V$, the control algorithm will favor minimizing the objective function $f_{obj}$ over the stability of the average rate queues. Our ultimate objective now is to minimize the following \textit{Lyapunov-drift-plus-penalty} function:
\begin{equation}\label{opt}
   g(t)=  \Delta(\Theta(t)) + V.\mathbb{E} \{f_{obj} \mid\Theta(t) \}
\end{equation}
Its upper bound, $\gamma(t)$, can be derived as follows for any action, any possible value of $\Theta(t)$ and any parameter $V>0$ \cite{neelyStochasticNetworkOptimization2010}:
\begin{equation}\label{ldpp}
\begin{split} 
        \gamma(t)  =& V.\E \{f_{obj} \mid \Theta(t) \} + \E \{Z(t).(\overline{\zeta} - \zeta _o) \mid \Theta(t) \}\\
        & + B + \sum _{i=1}^2 Q_i(t).\E \{A_i(t) - b_i(t) \mid \Theta(t) \}
        \end{split}
\end{equation}

where B is a constant that satisfies:
\begin{equation}
    \begin{split}
        B \geq &\frac{1}{2} \sum _{i=1}^2 \E \{A_i^2(t) - b_i^2(t)\mid \Theta(t)\} \\
        & + \frac{1}{2}\E \{(\overline{\zeta} - \zeta _o)\mid \Theta(t)\} \\
        & - \sum _{i=1}^2 \E \{A_i(t).\min \{ Q_i(t),b_i(t)\}\mid \Theta(t)\}
    \end{split}
\end{equation}

and $b_i(t)$ is the quantity of $TB_n$ at time slot t that the queue $i$ can process.
\begin{equation}
    b_i(t) =\left\{\begin{matrix}
TB_n^{r_{n,1}} & \text{if $i=1$} \\
TB_n^{r_{n,c_j}}.1_{TB_n^{r_{n,c_j}}} & \text{if $i=2$ } \\
0& \text{otherwise}\\

\end{matrix}\right. 
\end{equation}
Under Slater's condition \cite{neelySimpleConvergenceTime2014}, the drift-plus-penalty algorithm provides a  time-averaged $\mathscr {O}(1/ \epsilon$) of the expected size on all actual queues.
\begin{equation}\label{slatter}
\begin{split}
    &\limsup_{t \to \infty}{\frac{1}{t}.\sum _{\tau =0} ^{t-1}\overline{Q_1(\tau)+Q_2(\tau)}} \\
    &\leq \frac{B+C+V[f_{obj,max}-f_{obj,min}]}{\epsilon}
\end{split}
\end{equation}
where $C,\epsilon>0$ are constant values.


According to Equations (\ref{opt}) and (\ref{slatter}), by increasing the value of $V$ minimizes $f_{obj}$ at a cost of higher average queue length $\overline{Q_1(t)+Q_2(t)}$ and vice versa.

Through the opportunistic minimization framework of a conditional expectation \cite{neelyStochasticNetworkOptimization2010}, by minimizing $\gamma(t)$, the upper bound of the dual objective optimization, we can guarantee that the optimization problem \ref{eq:P1} will be satisfied.
Therefore, the design of our algorithm will be based on the control action $a$ at the decisive time $t_a$ and will choose the control action that exhaustively minimizes function $\gamma(t)$ as follows: 
\begin{algorithm}
	\caption{Control algorithm} 
	\begin{algorithmic}[1]
	\State Observe time slot t
	\If {$t\neq t_a$} 
	\State $t=t+1$
	\Else
	\State Observe the concatenated queue $\Theta(t)$
	\State Choose optimal action : $a$ = $\underset{a}{\arg \min }\{\gamma(t)\}$
	\EndIf
	\State Observe the outcomes of taken action 
	\State Update the queue $\Theta(t+1)$
	\end{algorithmic} 
\end{algorithm}

Since the optimal action is chosen based on the exhaustive search minimizing the designed function $\gamma(t)$ and the action space is bounded from $r_{min}$ to $r_{max}$, the computation complexity is relatively low and does not lead to an increase in the transmission processing time.

\section{Results and Discussions}
In this section, we compare the performance of several HARQ strategies (i.e. our proactive adaptation algorithm, a fixed proactive and a reactive) between a gNB and single mobile user. Initially, the UE is placed at a distance $d_0$ from the gNb and moves away from it with constant speed and direction ($v_x$, $v_y$). In this work, packets are generated in distinct ON and OFF periods that follow the Internet Protocol (IP) traffic model \cite{marviUseTrafficModels2019b}. 
The average duration of the ON and OFF periods are $t_{ON}$ and $t_{OFF}$, respectively. In the ON state, packets of variable size are generated with an arrival rate of $\lambda_{ON}$ and fill $Q_1(t)$.
The performance is evaluated in terms of RAN latency, reliability outage, packet loss and resource efficiency. 
We define the resource efficiency as the ratio of the number of radio resources required for a TB to be received by the receiver to the number of radio resources allocated by the scheduler. We also define the RAN latency as the times between the arrival of IP packets in the RLC layer of the gNb and their arrival in the  IP layer at the UE side. In the scheduling process, $K_1$ and $L_{12}$ \cite{3gppMediumAccessControl2019}  are modelled to illustrate the feedback processing time and data preparation time at the UE and gNB, respectively. For simplicity, we assumed that the core network latency and propagation delay are negligible. 
Table \ref{tab:sim_params} summarizes all parameters including communication band, transmit power $P_{tx}$, target Block Error Rate (BLER) $\epsilon_t$, maximum number of proactive cluster $c_{max}$ and proactive retransmission per cluster $r_{max}$ and risk threshold $\zeta_o$.


\begin{table}[ht]
\caption{Simulation Parameters}
    \centering
    \begin{tabular}{|p{3.3cm}|p{3cm}| }
    \hline
      \textbf{Parameters} &	\textbf{Values} \\
      \hline
       Packet Size &	Exponential(50) Bytes\\
       \hline
       ON-OFF Traffic &	$t_{on}$ = $t_{off}$ = 2.5 ms\\
       \hline
       Velocity ($v_x$,$v_y$) &	(4,4) $m/s$\\
       \hline
       $d_{0}$ &	110m\\
       \hline
        $(K_1,L_{12})$ &	(2,2) time slots\\
       \hline
       $R_{max}$ &	10\\
       \hline
       $(f_c,BW)$ &	(3.5 GHz, 50 MHz)\\
       \hline
       $P_{tx}$ &	8 dBm\\
       \hline
       $Numerology$ &	1\\
       \hline
       $(U_{tx}, S_{rx})$ &	($4\times 4, 2\times 2$)\\
       \hline
       $BLER$  $\epsilon_t$ &	$10^{-4}$\\
       \hline
       ($m , \eta_{S.eff} , CR$)  &	(5 , 0.7402 , 0.3701) \\
       \hline
       $c_{max}$ &	5\\
        \hline
       $(r_{min}, r_{max})$ &	(2, 5)\\
        \hline
        $\zeta_{o}$ & 0.05\\
        \hline 
    \end{tabular}
    \label{tab:sim_params}
 \end{table}

\begin{figure}[htp!]
\centering
\includegraphics[width=0.4\textwidth]{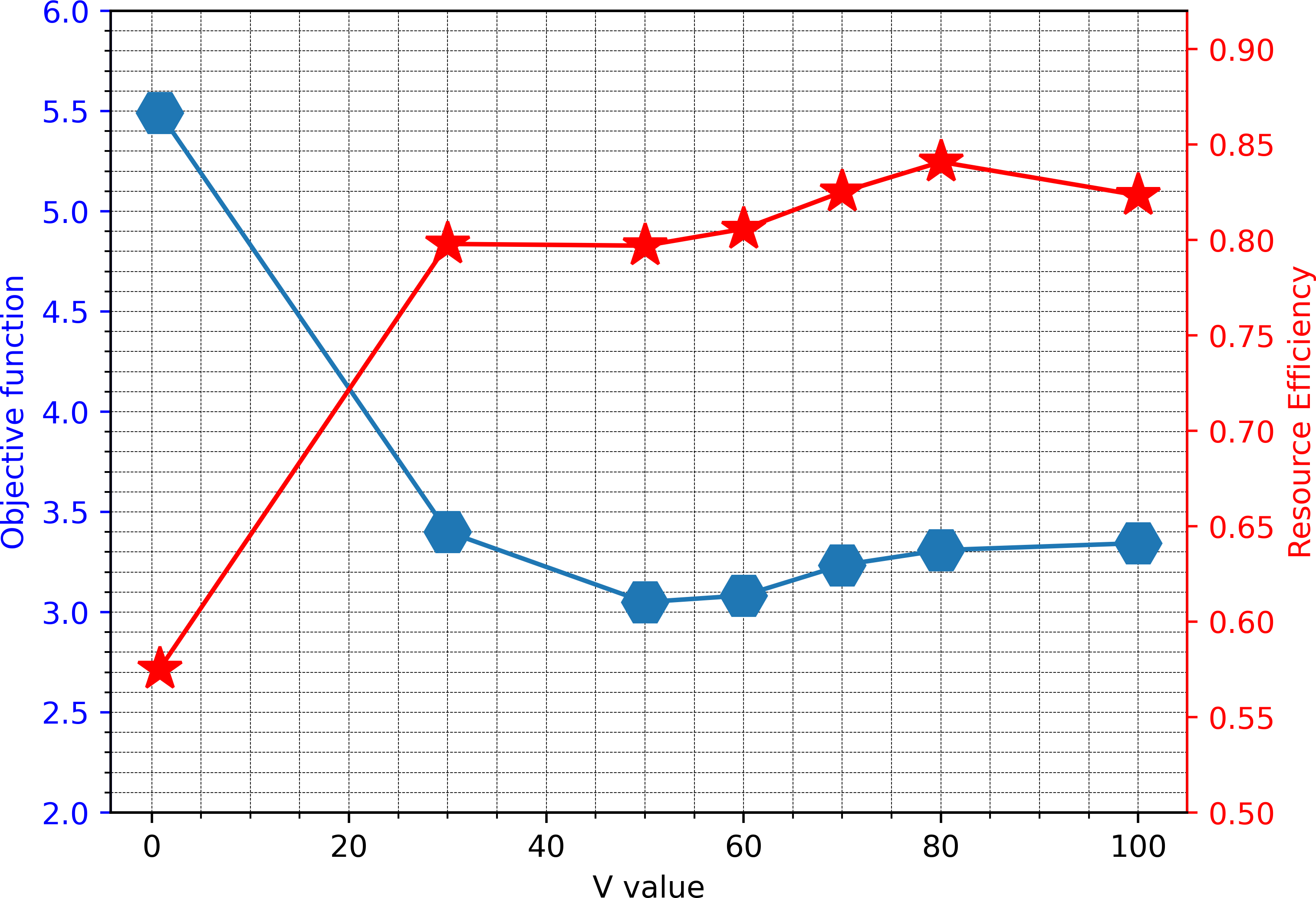}
\caption{Objective function and resource efficiency as a function of V}
\label{fig:objV}
\end{figure}

Figure \ref{fig:objV} illustrates how the objective function and resource efficiency behave as a function of the parameter V. For small value of V, our algorithm tends to minimize the expected changes in the Lyapunov function, i.e $L(\theta(t))$ rather than the number of allocated resources i.e. $f_{obj}$. By doing so, a large number of radio resources are generously provided to each TB for proactive RTXs, so the average resource allocation is high and the resource efficiency is low. When the value of V increases, the focus is on minimizing the objective function and fewer resources are allocated. At a certain value of V (i.e. around 50), we found that a minimum value of the objective function is reached (i.e. around 3.2). 
This is because at a high value of V, the algorithm favors allocating smaller number of parallel RTXs on each cluster and thus more proactive RTX clusters are required for each TB. 
Therefore, in our future evaluations, we will set V to 60.

\begin{figure}[htp!]
\centering
\includegraphics[width=0.46\textwidth]{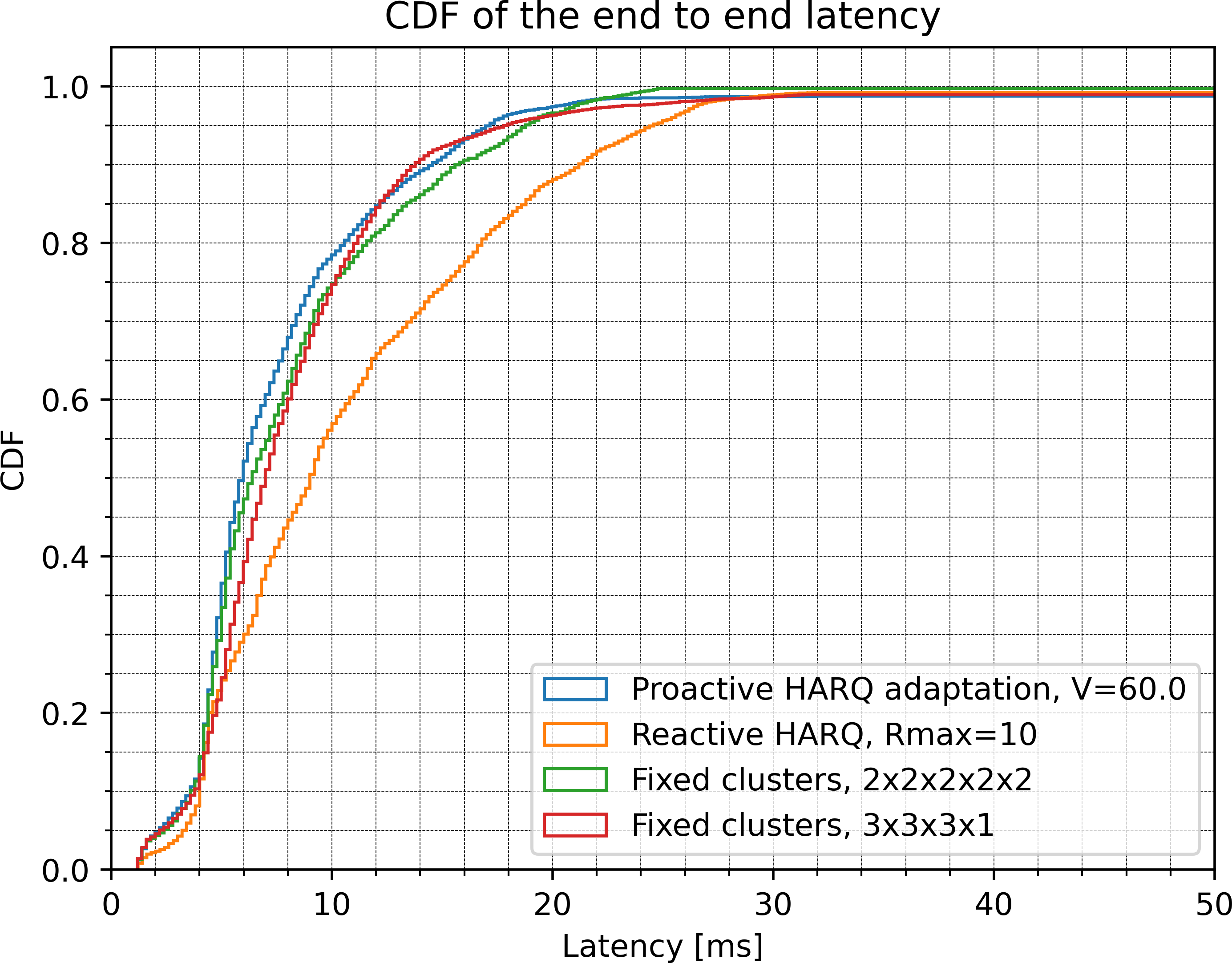}
\caption{CDF of latency for reactive RTX, fixed 2-parallel RTX, fixed 3-parallel RTXs and our dynamic adaptation algorithm with V=60}
\label{fig:cdf}
\end{figure}

Figure \ref{fig:cdf} compares the Complementary Distribution Function (CDF) of latency for different HARQ schemes: \textit{(i)} reactive HARQ scheme (i.e. $R=1$) with a maximum number of 10 RTXs, \textit{(ii)} 5 RTX clusters where each cluster contains 2 proactive RTXs (2-2-2-2-2 clusters), \textit{(iii)} 4 RTX clusters where the first three clusters contain 3 proactive RTXs and the last cluster contains a single RTX (3-3-3-1 clusters) and \textit{(iv)} our proposition with a maximum of 5 clusters. Figure \ref{fig:cdf} shows that the reactive HARQ scheme performs the worst compared to the other solutions due to the high RTT cost associated with triggering many RTXs. By enabling two parallel RTXs per received NACK, the latency can be improved at the cost of decreasing resource efficiency to around 0.82 compared to 1 in the reactive case. The latency can be further improved with 3 proactive RTXs per cluster, but the resource efficiency drops significantly to 0.69 when we redundantly provide radio resources for a TB to be successfully decoded. 
Concerning the performance of our algorithm with the adaptation of proactive HARQ, the performance of latency is slightly better than the others cases. As the channel condition and traffic behaviour varies dynamically, our dynamic decision maker will wisely select different redundancy level according to the current queues status and instantaneous channel condition. As the result, we can enhance the latency while keeping a good level of resource efficiency at around 0.8.

\begin{figure}[htp!]
\centering
\includegraphics[width=0.48\textwidth]{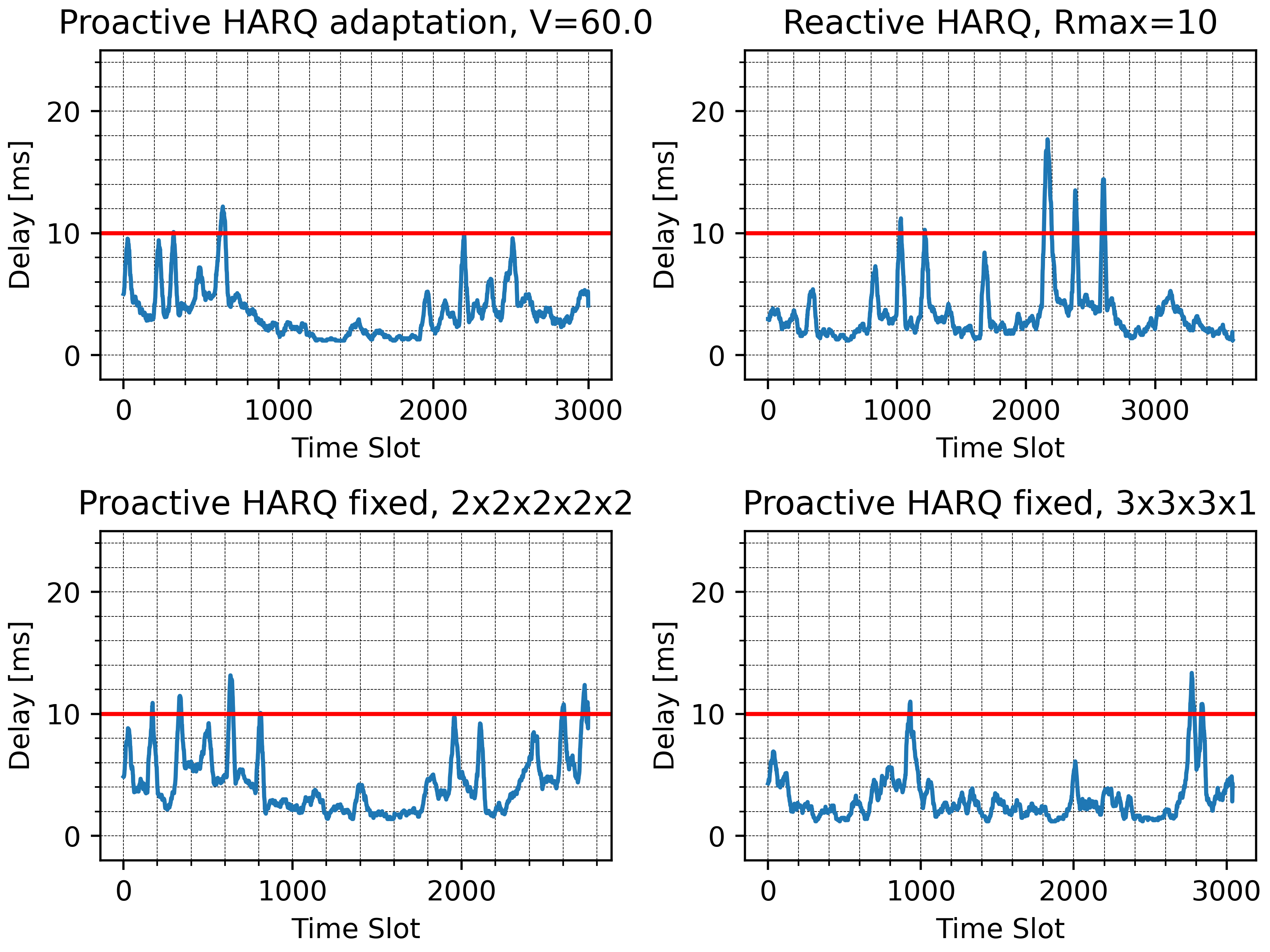}
\caption{Evolution of MAC delay between HARQ schemes}
\label{fig:delay}
\end{figure}

To show the temporal variation in MAC layer delay of the different HARQ schemes for intermittent URLLC traffic scenarios under time-varying channel conditions, Figure \ref{fig:delay} compares the latency between the departure of a TB at the sender's MAC layer and its successful arrival at the receiver's MAC layer. The reactive HARQ scheme experiences more peaks as it needs more time to successfully decode a packet when critical errors occur. 
When fixed proactive HARQ schemes are applied, 2-parallel strategy 
reduces the peak delay and 3-parallel strategy 
further improves the delay when the radio resources allocated in parallel help decode the packet faster. Our proposed algorithm also reduces the number and amplitudes of peaks when it dynamically selects an optimal level of proactive redundancy for each cluster, thus achieving a better trade-off between reserving more radio resources to recover the corrupted packet faster and less to maintain a minimum objective function.


\begin{figure}[htp!]
\centering
\includegraphics[width=0.4\textwidth]{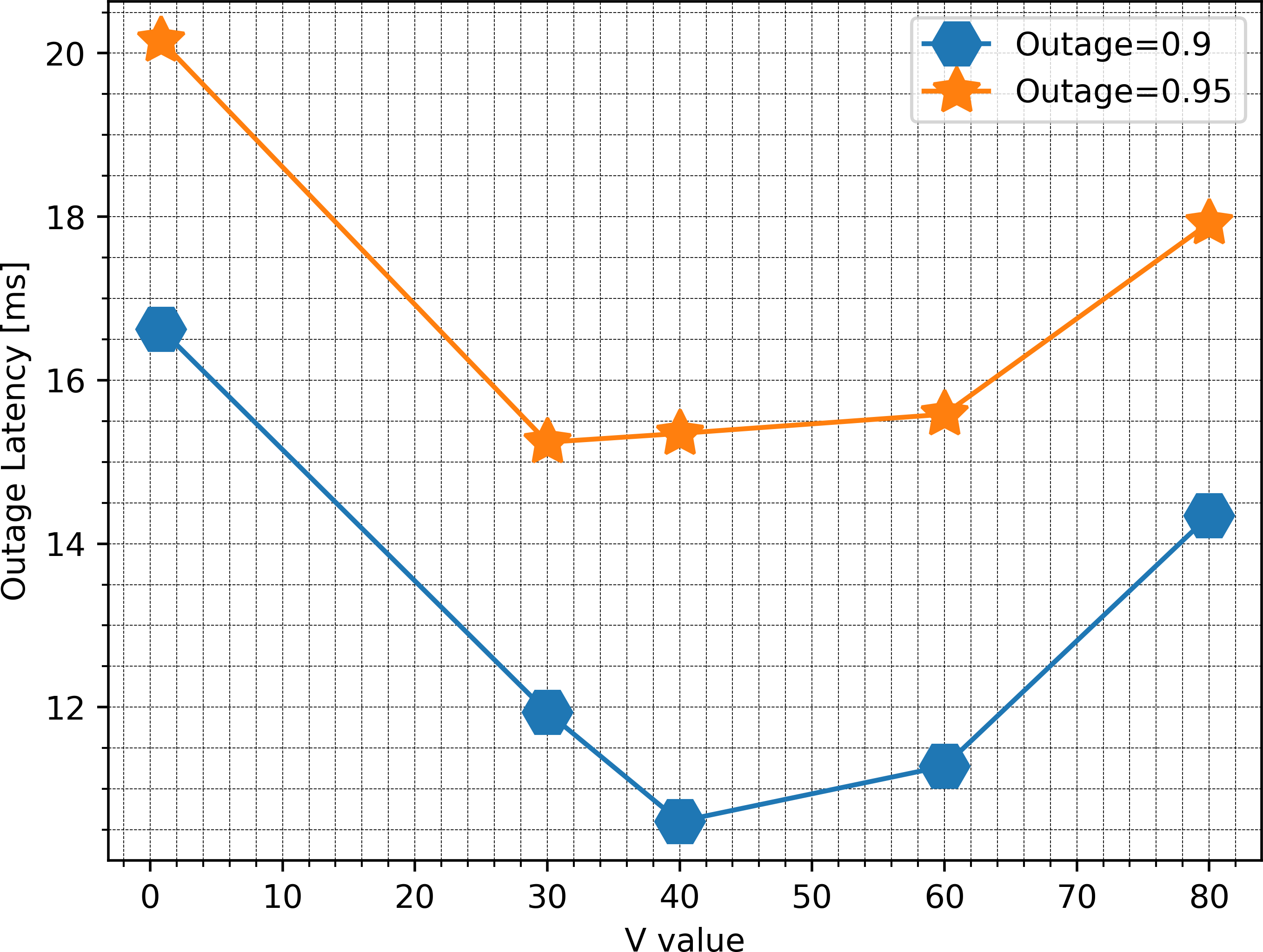}
\caption{Outage latency as the function of V value}
\label{fig:outage}
\end{figure}

Figure \ref{fig:outage} shows the latency as a function of V value for an outage level of 0.9 and 0.95, respectively. Certain values of V make the outage latency minimal because they allow a better balance between the number of resources allocated for each packet and the stability of the queues. As the result, the effect of overloaded buffer is better managed and leads to a good latency outage. On the other hand, by setting V too high or too low, this balance is no longer efficiently controlled and it causes a queuing effect due to the fact that too many clusters are used (large V) or too many retransmissions in each cluster are set (small V).




In addition, various performance metrics such as application packet loss (APP loss), resource efficiency as well as average  delay and its standard deviation for different HARQ schemes are discussed. First, the APP loss is guaranteed to be between 0.8 $\%$ and 1 $\%$ because the maximum number of retransmissions (i.e. $R_{max}=10$) is applied in all cases. Second, we find the largest average delay and its standard deviation in reactive HARQ whose values are around 10.77 ms and 6.57 ms, respectively. These coupled values are enhanced to roughly 8.14 ms, 4.91 ms and 8.1 ms, 4.5 ms in fixed 2-parallel and 3 parallel HARQ schemes at the expense of resource efficiency which decreases to 0.82 and 0.69, respectively. By dynamically deciding the number of proactive retransmissions at V=60, the average delay and jitter are slightly better than other HARQs which are about 7.2 ms and 4 ms while maintaining a good resource utilization level at about 0.8.  


\section{Conclusions}

In this study, we demonstrate the application of dynamic decision maker framework that enables a novel proactive HARQ design to cope with a time-varying channel and intermittent traffic source rate. 
Based on Lyapunov stochastic optimization tool, a mathematical framework is proposed to understand the performance-delay trade-off by minimizing the objective function of the total resource allocation and the total queue length that is parameterized by a V value. 
Our results reveal that an appropriate selection of V  enables the dynamic selection of proactive retransmission to overcome the defect of a long RTT in the reactive scheme while maintaining a good level of resource efficiency that is considered a drawback of the fixed proactive schemes.

In future work, this study will be extended to dynamic resource scheduling. The resource efficiency-latency-reliability trade-off will be achieved by the number, timing and intensity of the decisions. Moreover, we will implement our optimization for real experimentation with OpenAirInterface.

\section*{Acknowledgment}
This work was supported by the European Union H2020 / Taiwan Project 5G CONNI \cite{5GCONNI} under grant n.861459.


\bibliographystyle{ieeetr}
\bibliography{./Biblio}

\end{document}